\documentclass[cits]{PoS}

\newcommand{\nn}{\nonumber}

\title{Contribution from Duality Violations to the theoretical error on $\alpha_s$ }

\ShortTitle{Duality Violations}

\author{Oscar Cat\`a\\
       INFN, Laboratori Nazionali di Frascati, Via E. Fermi 40, I-00044 Frascati, Italy\\
       E-mail: \email{ocata@ifae.es}}

       \author{Maarten Golterman\\
     Department of Physics and Astronomy, San Francisco State
University, 1600 Holloway Ave, San Francisco, CA 94132, USA\\
       E-mail: \email{maarten@stars.sfsu.edu}}

\author{\speaker{Santiago Peris}\thanks{This work was supported in part by
CICYT-FEDER-FPA2008-01430, SGR2005-00916, the Spanish Consolider-Ingenio 2010 Program CPAN (CSD2007-00042), by the EU Contract No. MRTN-CT-2006-035482, ``FLAVIAnet,'' and by the US Department of Energy.
}\\
        Grup de F{\'\i}sica Te{\`o}rica and IFAE, UAB, E-08193 Bellaterra, Barcelona, Spain\\
        E-mail: \email{peris@ifae.es}}

\abstract{Duality Violations (DVs) is a nickname for the failure of the Operator Product Expansion to describe QCD correlators on the physical axis.  Using a physically motivated ansatz, a fit to the spectral functions allows us to get a quantitative estimate for the amount of DVs present in $\tau$ data. The quality of the fit turns out to be better than expected. Since DVs have not been included in the past in the determination of $\alpha_s$, they amount to an additional  theoretical error which we estimate could be  $\delta \alpha_s(m_{\tau}) \sim  0.003-0.010 $. Our ansatz satisfies, in particular, the 1st Weinberg sum rule,  which shows  that this sum rule is not enough to force  DVs to vanish.}

\FullConference{International Workshop on Effective Field Theories: from the pion to the upsilon \\
		February 2-6 2009\\
		Valencia, Spain}

\begin{document}

\section{Introduction}

Our knowledge of $\alpha_s$ is of fundamental importance. This statement, which is obviously true in QCD, can potentially go far beyond the Standard Model. Let us remember, for instance, that the value of $\alpha_s(M_Z)$ strongly affects the physics at much higher scales in the context of Grand Unification\cite{Raby}. An increasingly accurate determination of $\alpha_s$ is, therefore, an important  goal which fully deserves the continuous effort it has attracted \cite{Hinchliffe}.

When determining $\alpha_s$, the decay of the tau lepton has proved to be a very convenient process \cite{BNP}. For one thing, it is the only lepton which is heavy enough to decay into hadrons, so its decay obviously happens without the complications of a hadronic contamination in the initial state. Furthermore, thanks to a tremendous experimental effort,  there are also very accurate data available \cite{Aleph}.

The flip side of all the previous advantages is that the tau mass, $m_{\tau}\simeq 1.77\ \mathrm{GeV}$,  is very close to the QCD scale, $\Lambda_{\mathrm{QCD}}\sim 1\ \mathrm{GeV}$, which means that effects going beyond  perturbation theory  have to be brought under very good theoretical control in order to be able to reliably obtain a very accurate determination of $\alpha_s$. Or else, discrepancies arise. For instance, recent determinations of $\alpha_s$ find the following values:
\begin{eqnarray}\label{alphas}
\alpha_s(m_{\tau}^2)&=0.344\pm 0.005_{\mathrm{exp}}\pm 0.007_{\mathrm{th}}\ [5] \ , \quad \alpha_s(m_{\tau}^2)&=0.332\pm 0.005_{\mathrm{exp}}\pm 0.015_{\mathrm{th}}\ [6] \ , \nonumber \\
\alpha_s(m_{\tau}^2)&=0.321\pm 0.005_{\mathrm{exp}}\pm 0.012_{\mathrm{th}}\ [7]\ , \quad
\alpha_s(m_{\tau}^2)&=0.316\pm 0.003_{\mathrm{exp}}\pm 0.005_{\mathrm{th}}\ [8]\  .
\end{eqnarray}
With the high level of precision claimed, these determinations are not compatible. So, either there are assumptions which went into the corresponding analysis which are not correct, or the errors have been underestimated.

 Perturbative effects (even after resummations) are not all there is. After all hadrons cannot be obtained solely from summing Feynman diagrams and there are also genuinely nonperturbative effects. These are understood mostly in the framework of the Operator Product Expansion (OPE), where they take the form of condensates \cite{SVZ}\cite{condensates:theworks}, but there may be others\cite{Narison}. In this framework there is a potential problem since the OPE is supposed to be a valid expansion only in the euclidean (i.e. $Q^2>0$), whereas the experimental data are fully restricted to the minkwoski region (i.e. $Q^2<0$). In order to connect the two regions  an assumption about analytic continuation has to be made.  For inclusive quantities like tau decay, it amounts to using Cauchy's theorem on the OPE instead of using it on the full Green's function, which is the one truly satisfying the right analytic properties. This assumption, first proposed in Ref. \cite{Shankar}, has been employed in all analyses of tau decay up to date, starting with the pioneering work in Ref. \cite{BNP}. The amount by which this assumption fails is referred to as Duality Violations (DVs) (see Eq. (\ref{cauchy3}) and below  for a more precise definition). Obviously, if this  assumption is not right with enough accuracy, an associated theoretical error should be included  in the final error for $\alpha_s$.

It is clear that the properties of the OPE are crucial for understanding the physics of tau decay. For instance, a recent analysis \cite{Davier}  based on Aleph data finds for the gluon condensate the following values:
\begin{equation} \label{gluonVA}
\frac{\alpha_s}{\pi}\langle GG\rangle\Big|_{\mathrm{Vector}}\!\!\!\!\!\!\!\!\!\!\!\!=(-0.8\pm0.4)\times
10^{-2}\ {\mathrm{GeV}}^4\ , \quad
\frac{\alpha_s}{\pi}\langle
GG\rangle\Big|_{\mathrm{Axial}}\!\!\!\!\!\!\!\!\!\!\!=(-2.2\pm0.4)\times 10^{-2}\ {\mathrm{GeV}}^4\ ,\nn
\end{equation}
depending on whether the vector or the axial-vector spectral functions are employed. These two values are not compatible and, if taken literally,  they would signal a clear breakdown of the OPE.

Given the situation, and the precision sought, we think it is time for a reassessment of the total theoretical error involved. Studies have been made of the error involved in the different resummations of perturbation theory \cite{Jamin,Menke}, and also of the combination of spectral functions which may be optimal \cite{Maltman, Dominguez}, but the amount of work dedicated to assess the validity of the assumption that DVs are absent is surprisingly meager. This is so even though there are  rather general expectations that it must fail at some level \cite{Shifman}.  This is partly why we decided to take a fresh look at this problem and make a reasonable estimate of the theoretical error involved  \cite{Thebomb, CGP1, CGP2}.

Defining
\begin{equation}\label{dv1}
    \Delta_{V,A}(q^2)=\Pi_{V,A} (q^2)-\Pi^{ \mathrm{OPE}}_{V,A} (q^2)  \ ,
\end{equation}
the central equation for the discussion is given by \cite{CGP1,CGP2}
\begin{equation}\label{cauchy3}
\int_0^{s_0}\,ds\,\, P(s) \,\frac{1}{\pi}{\mathrm{Im}}\,\Pi_{V,A}(s)=-\frac{1}{2\pi i}\oint_{|q^2|=s_0}\,dq^2
\,P(q^2)\, \Pi^{OPE}_{V,A} (q^2)+ {\cal{D}}_{V,A}^{[P]}(s_0) \  .
\end{equation}
where
\begin{equation}\label{ourbaby}
    {\cal{D}}_{V,A}^{[P]}(s_0) =- \int_{s_0}^{\infty} ds\ P(s)\  \frac{1}{\pi} \mathrm{Im}\Delta_{V,A}(s)\ .
\end{equation}
is the function encoding all the DVs. In particular, if the OPE were a convergent expansion in the region of interest, one would have $ \Delta_{V,A}(q^2)=0$ and no duality violations. Effectively, this is the assumption made in all analyses up to now. The problem with this assumption is that, on the minkowski axis, where the data are located, this obviously cannot happen as the OPE does not reproduce the spectrum, i.e. we \emph{know} that  ${\mathrm{Im}} \Delta_{V,A}(q^2)\neq 0$. Therefore,  the question to address is rather how large $ {\cal{D}}_{V,A}^{[P]}(s_0)$ can be. Regretfully, and this probably explains why the question has not been addressed before, there is no theory of DVs so the answer cannot be obtained in QCD from first principles.

\section{Extracting Duality Violations from $\tau$ data}

\begin{figure}
\centering
\includegraphics[width=2.3in]{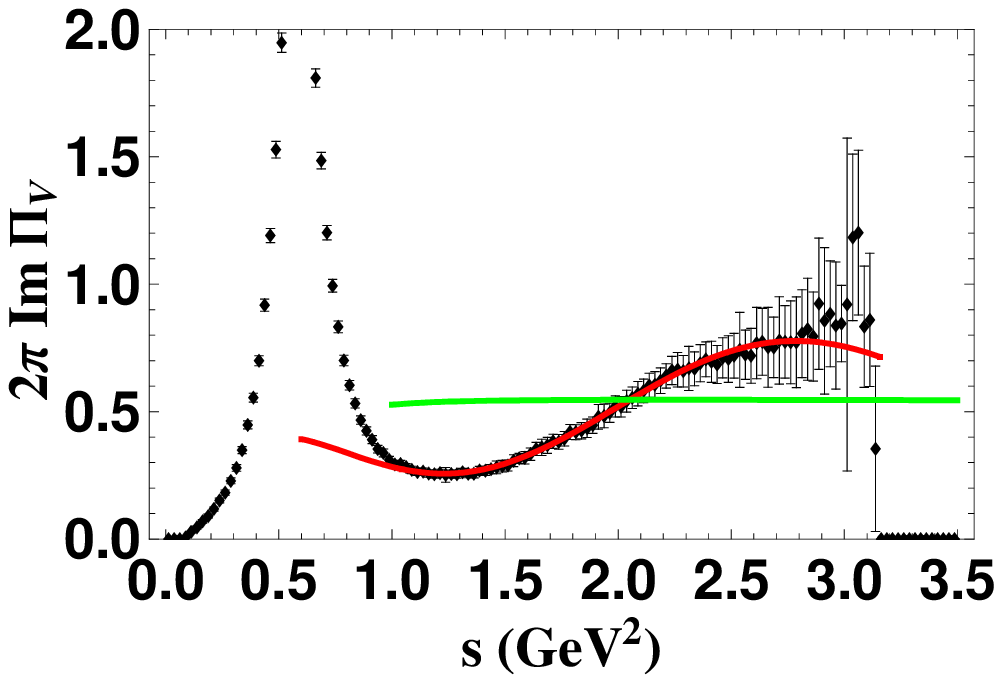}
\hspace{.2cm}
\includegraphics[width=2.3in]{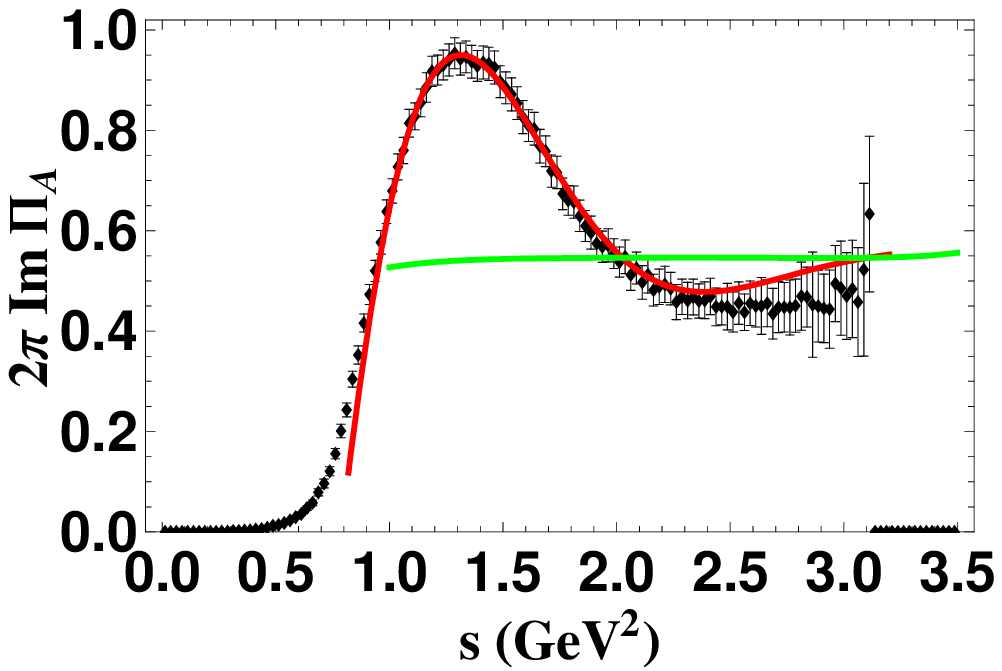}
\includegraphics[width=2.3in]{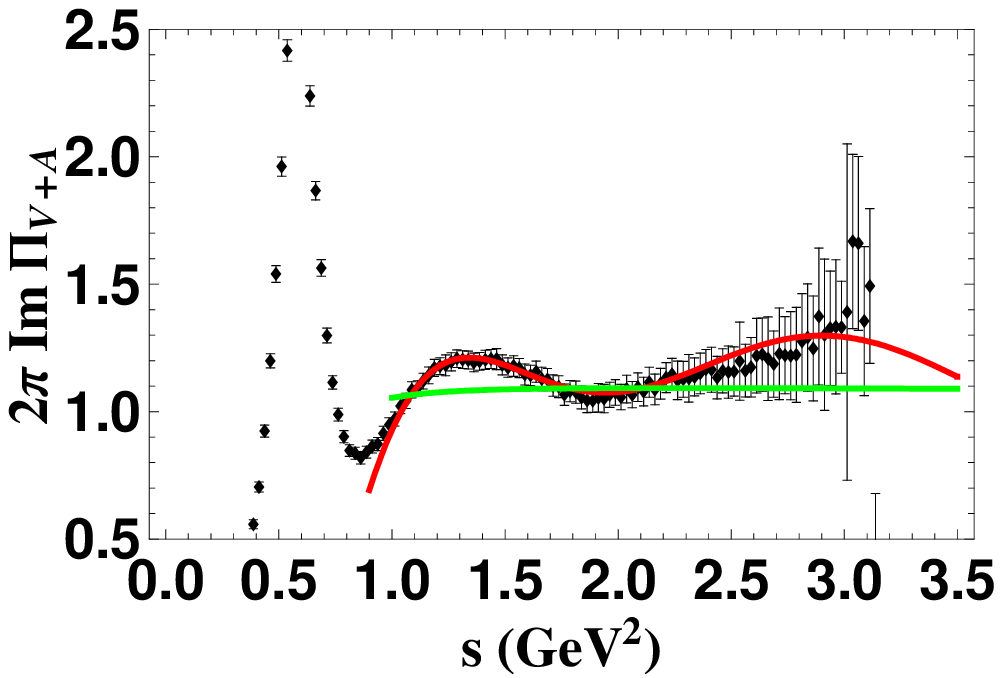}
\hspace{.2cm}
\includegraphics[width=2.3in]{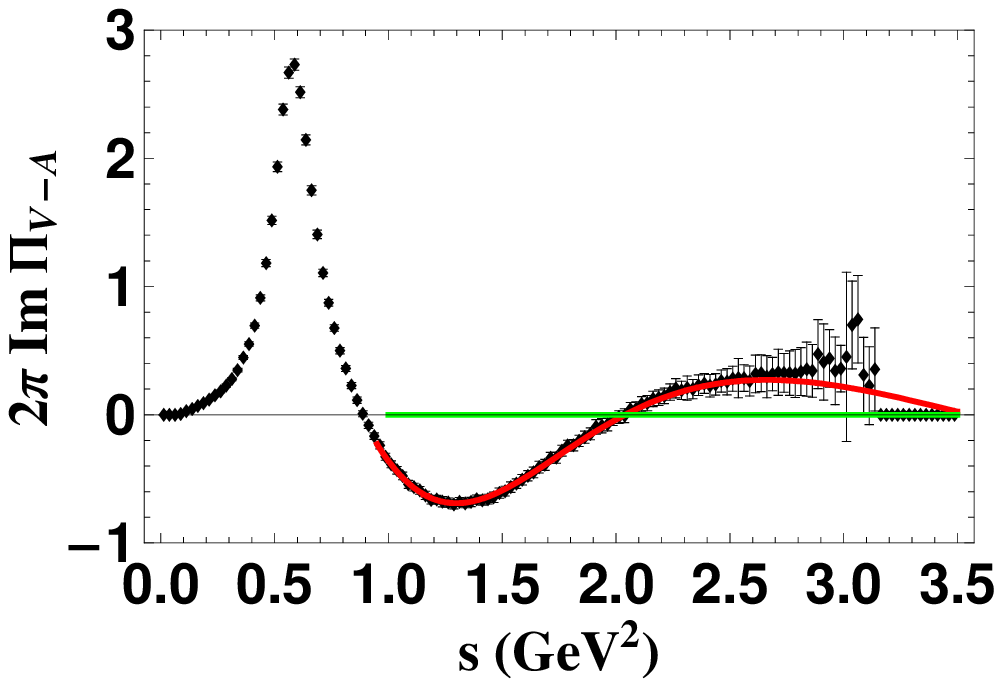}
\caption{Different combinations of spectral functions ($V$, $A$, and $V\pm A$)
(data points in black) compared with Fixed-Order perturbation theory (green flat line) and the result of our fit (2.1-2.2) (red curve).}\label{aleph-fig}
\end{figure}

The function $\mathrm{Im}\Delta_{V,A}(s)$ in Eq. (\ref{ourbaby}) requires knowledge of the spectrum up to infinite energy and this is, in principle, unknown. This is the main difficulty. In order to make progress, it is necessary to have a physically motivated model with which to extrapolate the data from the region below the tau mass to the region above it.

Motivated by very general arguments on Regge theory and asymptoticity of the OPE, we would like to suggest the following simple form for the parametrization of the DVs:
\begin{equation}\label{ansatz}
   \frac{1}{\pi} \mathrm{Im}\Delta_{V,A}(s)= \theta(s-s_{min})\  \kappa_{V,A} \ e^{-\gamma_{V,A} s}\ \sin\left(\alpha_{V,A} +\beta_{V,A} s \right)\ .
\end{equation}
The exponential fall-off is expected to originate from the intrinsic error inherent to an asymptotic expansion and the sine function from a harmonic expansion of the periodic function representing the daughter repetitions in the spectrum of Regge theory. The step function ensures the validity of the parametrization (\ref{ansatz}) only at high-enough energies. In other words, (\ref{ansatz}) represents the first correction to the asymptotic behavior of the spectral function at high energies given by the parton model plus condensates. In Refs. \cite{Shifman}-\cite{CGP2} a model is studied which realizes all the expected features known of the OPE in QCD and Regge theory, and for which Eq. (\ref{ansatz}) is the correct behavior.

A fit of perturbation theory\footnote{Condensates give negligible contributions \cite{Thebomb}.}plus the parametrization (\ref{ansatz}) to the spectral functions, in the window $1.1$ GeV$^2 \leq s \leq m_{\tau}^2$, yields \cite{Thebomb}:
\begin{eqnarray}\label{fitV}
   \kappa_V = 0.018\pm 0.004 \qquad \qquad &,& \quad  \kappa_A = \ 0.20\pm 0.06 \quad \quad \ , \nn \\
  \gamma_V = 0.15\pm 0.15  \quad  \mathrm{GeV}^{-2}\quad &,& \quad   \gamma_A = \  1.7\pm 0.2  \quad  \mathrm{GeV}^{-2}\ , \nn\\
   \alpha_V = 2.2\pm 0.3 \qquad \qquad \qquad &,& \quad    \alpha_A = -0.4\pm 0.1 \quad \quad , \nn \\
 \beta_V = 2.0\pm 0.1  \quad \mathrm{GeV}^{-2}\qquad &,& \quad  \beta_A = -3.0\pm 0.1 \quad\   \mathrm{GeV}^{-2}\ , \nn \\
   \frac{\chi^2}{dof} = \frac{10}{79}\simeq 0.13\quad \qquad \qquad &,& \quad  \frac{\chi^2}{dof} = \frac{17}{78}\simeq 0.22\quad .
\end{eqnarray}
The result of this fit is compared to the data in Fig. \ref{aleph-fig}. As one can see from the plots and the  corresponding $\chi^2/dof$, the quality of the fits  is more than acceptable.

\begin{figure}
\centering
\includegraphics[width=2.8in]{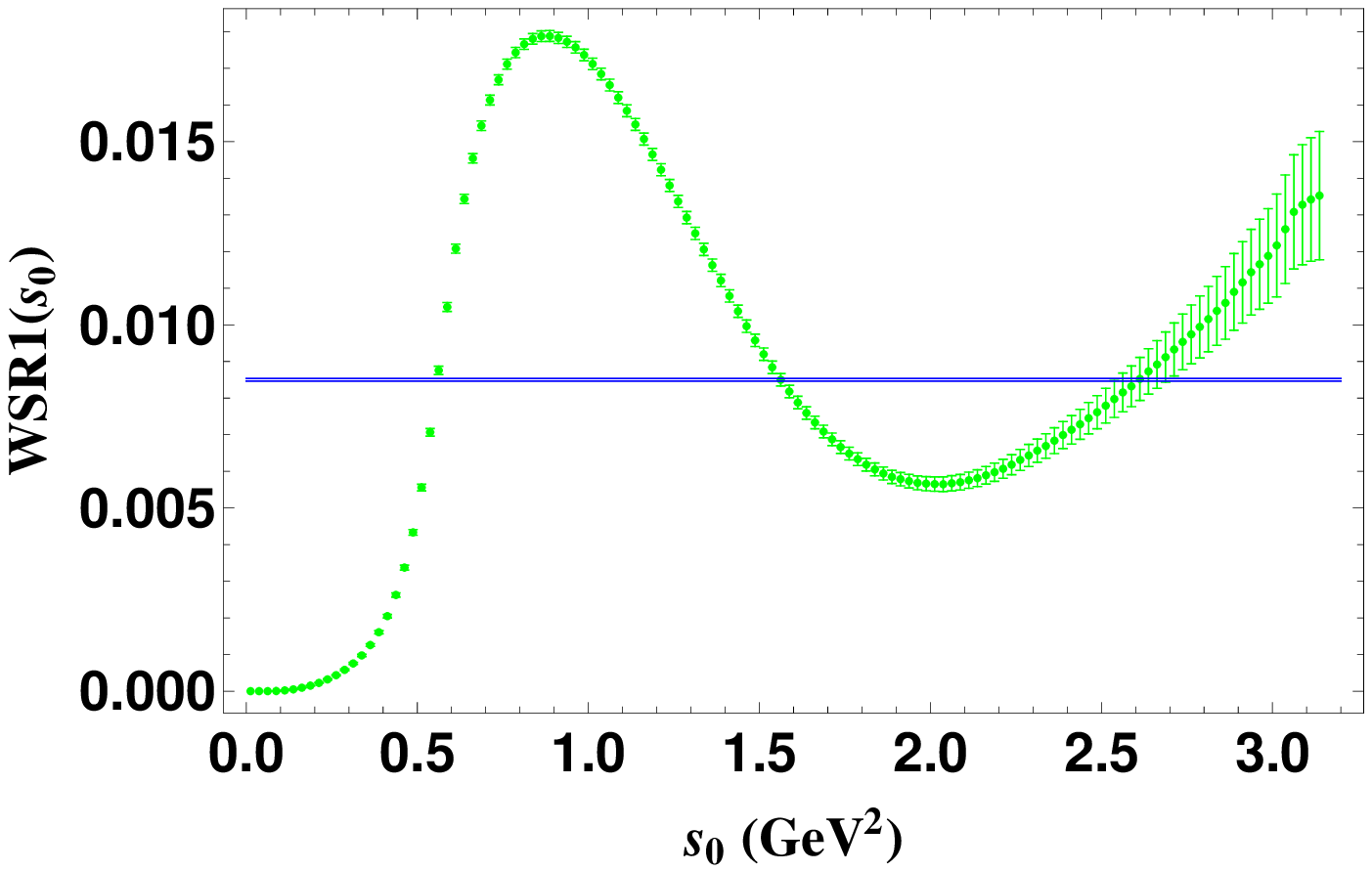}
\hspace{.5 cm}
\includegraphics[width=2.8in]{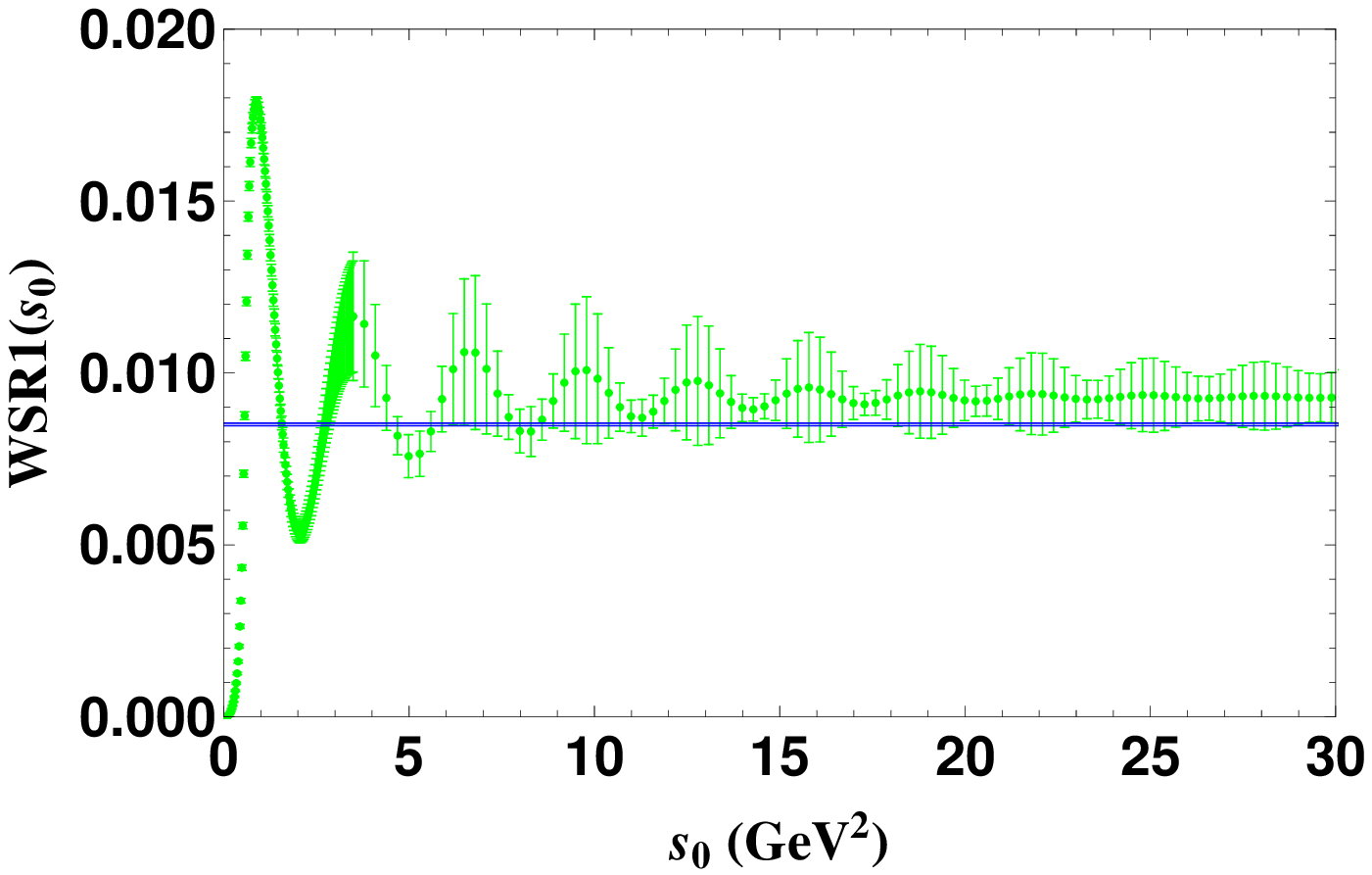}
\caption{Left panel: result of the WSR1 for $s_0=m_{\tau}^2$ employing $\tau$ data. Right panel: the WSR1 extrapolated to higher energies with the use of our ansatz (2.1-2.2). The (blue) horizontal line is given by the value of $f^2_{\pi}$.}\label{WSR}
\end{figure}

Since now one has a way to extrapolate the spectral function all the way to infinity, it makes sense to ask whether the ansatz (\ref{ansatz}) together with the values for the parameters (\ref{fitV}) satisfies the 1st Weinberg sum rule\footnote{We remark that the 2nd Weinberg sum rule breaks down away from the chiral limit.} (WSR1).\footnote{We thank A. Pich for asking this question.} One may not expect, however, that the condition to satisfy WSR1 will do away with  DVs altogether. The reason is that $\alpha_s$ is determined from the $V+A$ combination of spectral functions in $\tau$ decay whereas WSR1 obviously depends on the orthogonal combination $V-A$ which contains no perturbative contribution (and, therefore, no dependence on $\alpha_s$). At any rate, due to the smallness of the $u,d$ quark masses, the WSR1 may be expressed,  within a very good approximation, as
\begin{equation}\label{WSR1}
    \mathrm{WSR1}(s_0)=\frac{1}{2 \pi}\ \int_0^{s_0}\ ds \
     \mathrm{Im}\Pi_{V-A} =f_{\pi}^2 \quad  .
\end{equation}
This sum rule, which is supposed to be valid only for $s_0$ very large, follows from the general result in Eq. (\ref{cauchy3}), \emph{provided} the DV contribution ${\cal D}_{V-A}^{[P=1]}(s_0)$ is set to zero.\footnote{Contributions from condensates are expected to be negligible.} This can be seen by taking the particular polynomial $P(q^2)=1$ and remembering that there is no condensate of dimension two from the OPE in the $V-A$ combination.  In other words , for any finite $s_0$, the WSR1 is a measure of DVs at that scale since only when $s_0\rightarrow \infty$ can one make sure that DVs vanish (see Eq. (\ref{ourbaby})).   The result of the integral up to an scale $s_0$ is shown in Fig. \ref{WSR}. On the left panel, one sees that the experimental data grossly violates the WSR1 at $s=m_{\tau}^2$,  exposing thereby the existence of DVs at this scale. At higher energies, the assumption of no DVs  is tantamount to essentially a constant zero line from $m_{\tau}^2$ onwards  for the $V-A$ combination of spectral functions shown on the rightmost lower panel in Fig. \ref{aleph-fig}.\footnote{Again, the contributions from the condensates are negligible \cite{Thebomb}.} This means that the result obtained on the left panel of Fig. \ref{WSR} for $s_0=m_{\tau}^2$ cannot improve at higher energies but remains constant. Our conclusion, therefore, is that DVs are clearly not vanishing at $m_{\tau}^2$.

On the contrary, when the data are extrapolated at higher energies taking into account DVs with our parametrization (\ref{ansatz},\ref{fitV}), the sum rule does get satisfied within errors, as it should. This is shown on the right panel of Fig. \ref{WSR}.

The presence of the DV term (\ref{ansatz}) with the values for the parameters extracted from the fit (\ref{fitV}) affects the standard extraction of $\alpha_s$ made through Eq. (\ref{cauchy3}) because of the contribution coming from ${\cal D}_{V,A}(s_0)$. We have estimated in Ref. \cite{Thebomb} the associated theoretical error in $\alpha_s$ due to DVs as
\begin{equation}\label{shift}
  \delta \alpha_s(m_{\tau}) \sim  0.003-0.010 \ ,
 \end{equation}
where the spread of values includes the propagation of all the errors involved. Another attempt at estimating this theoretical error was made in Ref. \cite{Davier}, with the result that DVs were negligibly small. However, unlike in our analysis, no detailed fits to the spectral data were made in this reference.

\section{Inclusion of $e^{+}e^{-}$ data. Conclusions}

\begin{figure}
\centering
\includegraphics[width=2.3in]{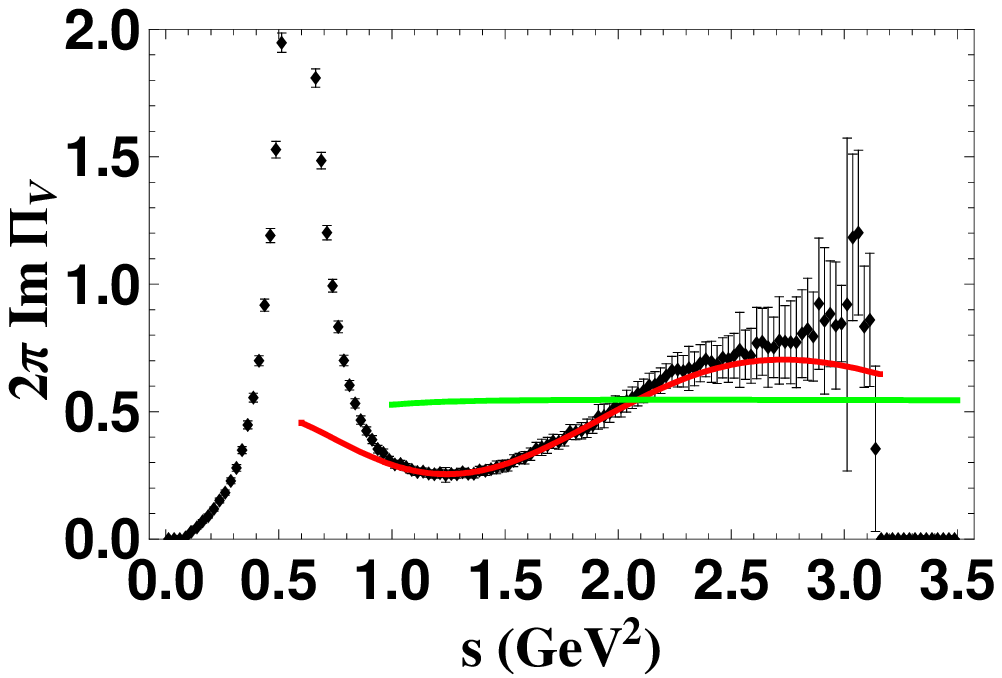}
\hspace{0.1cm}
\includegraphics[width=2.3in]{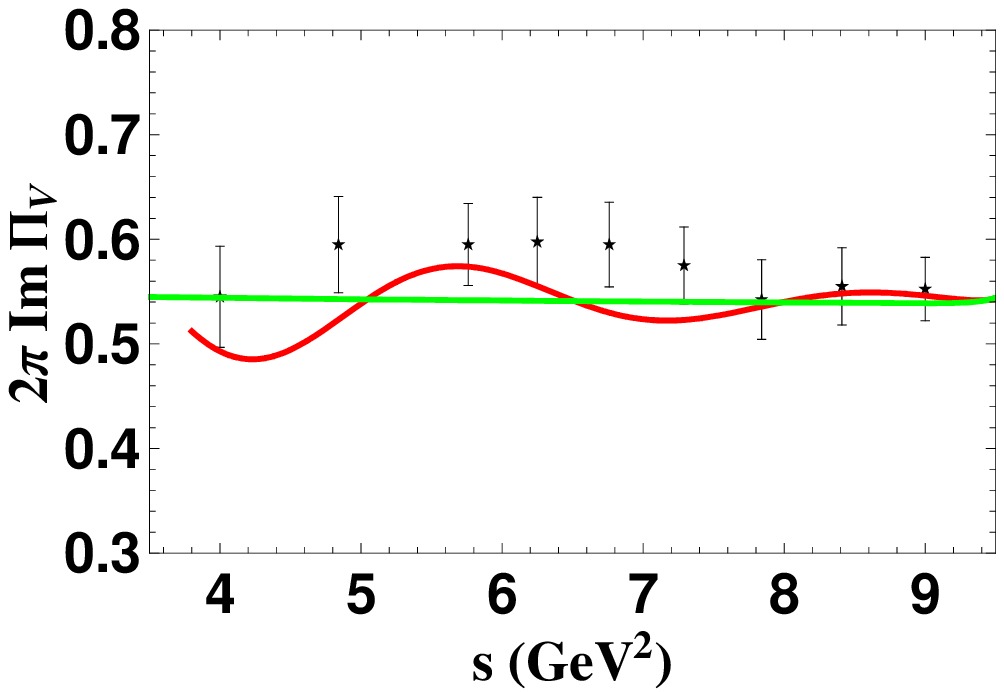}
\caption{Plots of the vector spectral function in tau decay (left panel) and $e^+e^-$ (right panel) as compared to perturbation theory (green flat line) and the result of the combined fit (3.1) (red oscillating curve). }\label{epem-fig}
\end{figure}

Since our DV ansatz allows one to go to higher energies, one may wonder how well our vector spectral function compares with the data extracted from $e^{+}e^{-}$.\footnote{Regretfully, there are no data at higher energies in the axial channel.} Of course, our ansatz (\ref{ansatz}) cannot be directly used for $e^+e^-$ data because it contains only isospin one (i.e. the $ud$ combination) whereas in $e^+e^-$ one has the flavor singlet combination $(2/3) \overline{u}u- (1/3) \overline{d}d - (1/3) \overline{s}s$. Neglecting this fact  and adjusting only for the different charges involved (a bold step to take) plus a shift in the parameter $\alpha \rightarrow \alpha'$ (to take into account that the $s$ quark is much heavier than the $u,d$; a modification which is suggested by the model in \cite{Shifman}-\cite{CGP2}) we have performed a simultaneous fit to the vector spectral function from $\tau$ decay, together with the vector spectral function in $e^+e^-$ above $4\ \mathrm{GeV}^2$ up to the charm threshold \cite{Bai}. The $e^+e^-$ data between $m_{\tau}^2$ and $4\ \mathrm{GeV}^2$ is controversial \cite{Eidelman} and we have not used it.

The result of this simultaneous fit becomes
\begin{eqnarray}\label{hybridfit}
  \kappa_V &=& 0.024\pm 0.004 \nn \\
  \gamma_V &=& 0.40 \pm 0.12 \nn \\
  \alpha_V &=& 1.82 \pm 0.19\nn \\
  \beta_V &=& 2.14\pm 0.11\nn \\
  \alpha'_V &=& 5.2\pm 1.4 \nn \\
  \frac{\chi^2}{dof} &=& \frac{22}{87}\simeq 0.25  \ ,
\end{eqnarray}
which entails a shift from the values obtained only with $\tau$ data in (\ref{fitV}), although it is compatible within errors. The result of the fit (\ref{hybridfit}) can be seen in Fig. \ref{epem-fig}. As one can see, given the assumptions made, our ansatz is not grossly incompatible with $e^+e^-$ data. Taking the values (\ref{hybridfit}) at face value, the theoretical error in $\alpha_s$ turns out to be smaller than (\ref{shift}) by a factor of $\sim 3$.

We think it is advisable not to fall on the optimistic side when it comes to estimating a  theoretical error. From the plots in Fig. \ref{aleph-fig} and \ref{WSR}, we conclude that a theoretical error in the determination of $\alpha_s$ from duality violations at the level of Eq. (\ref{shift}) is not at all excluded. However, this error has not been included up to now in any determination of $\alpha_s$ from $\tau$ decay. Until we learn more about DVs, we think it should. We refer to Ref.~\cite{Thebomb} for more details and further discussions.

\end{document}